\documentclass[
showpacs,superscriptaddress, aps, twocolumn, longbibliography]{revtex4-1} 

\usepackage[USenglish]{babel}

\usepackage{graphicx,xcolor}
\usepackage{amsmath, bbm}
\usepackage{natbib}
\usepackage{hyperref}
\usepackage{epstopdf}
\usepackage{color}

\renewcommand{\imath}[0]{\mathrm{i}}

\newcommand{\myref}[1]{}

\begin{document}
\title{On-chip linear and nonlinear control of single molecules coupled to a nanoguide}
 \author{Pierre T\"urschmann}%
\affiliation{Max Planck Institute for the Science of Light, Staudtstr. 2, D-91058 Erlangen, Germany}
 \author{Nir Rotenberg}%
\affiliation{Max Planck Institute for the Science of Light, Staudtstr. 2, D-91058 Erlangen, Germany}
\author{Jan Renger}%
\affiliation{Max Planck Institute for the Science of Light, Staudtstr. 2, D-91058 Erlangen, Germany}
\author{Irina Harder}%
\affiliation{Max Planck Institute for the Science of Light, Staudtstr. 2, D-91058 Erlangen, Germany}
\author{Olga Lohse}%
\affiliation{Max Planck Institute for the Science of Light, Staudtstr. 2, D-91058 Erlangen, Germany}
\author{Tobias Utikal}%
\affiliation{Max Planck Institute for the Science of Light, Staudtstr. 2, D-91058 Erlangen, Germany}
\author{Stephan G\"otzinger}%
\affiliation{Friedrich Alexander University Erlangen-Nuremberg, D-91058 Erlangen, Germany}
\affiliation{Max Planck Institute for the Science of Light, Staudtstr. 2, D-91058 Erlangen, Germany}
\author{Vahid Sandoghdar}
\affiliation{Max Planck Institute for the Science of Light, Staudtstr. 2, D-91058 Erlangen, Germany}
\affiliation{Friedrich Alexander University Erlangen-Nuremberg, D-91058 Erlangen, Germany}

\date{\today}

\begin{abstract}
While experiments with one or two quantum emitters have become routine in various laboratories, scalable platforms for efficient optical coupling of many quantum systems remain elusive. To address this issue, we report on chip-based systems made of one-dimensional subwavelength dielectric waveguides (nanoguides) and polycyclic aromatic hydrocarbon molecules. After discussing the design and fabrication requirements, we present data on coherent linear and nonlinear spectroscopy of single molecules coupled to a nanoguide mode. Our results show that external microelectrodes as well as optical beams can be used to switch the propagation of light in a nanoguide via the Stark effect and a nonlinear optical process, respectively. The presented nanoguide architecture paves the way for the investigation of many-body phenomena and polaritonic states and can be readily extended to more complex geometries for the realization of quantum integrated photonic circuits.
\end{abstract} 

\maketitle

In the past three decades, optical studies of single quantum systems have matured to become commonplace in many laboratories. A next grand challenge in quantum nano-optics will be to control mesoscopic assemblies of individual quantum systems, where only a few particles of light and matter interact, possibly in an entangled fashion. The very first steps in this direction, involving two quantum emitters have already been taken in various systems \cite{Maunz:07, Rezus:12, Ritter:12, Pfaff:14, Hood:16a, Sipahigil:16}, but the low efficiencies in such experiments hamper their scaling prospects. 

To achieve significant correlated dynamics and polaritonic effects, it is desirable to couple many quantum emitters with large scattering cross sections to a common spatial photonic mode, all at the same transition wavelength $\lambda$ \cite{Greentree:06, Carusotto:13, Haakh:16, noh:16}. One particularly attractive approach for realizing this scenario is to couple the emitters to a one-dimensional subwavelength waveguide (nanoguide), which can act as an optical bus for communication among emitters at distances much larger than $\lambda$. Several groups have recently taken pioneering steps in this direction by coupling atoms \cite{Vetsch:10, Goban:14}, semiconductor quantum dots \cite{Yalla:12, Arcari14}, and molecules \cite{Stiebeiner:09, Faez:14, Skoff:16} to nanoguides. Each system has some merits and confronts some challenges. In particular, reaching high densities for the realization of complex cooperative effects \cite{Carusotto:13, Haakh:16, noh:16} remains a nontrivial task. Here, organic molecules offer a unique advantage because they can be embedded in a solid matrix at densities of several thousands  per cubic wavelength \cite{SMbook}. 

\begin{figure}[t!]
\includegraphics[width=7.1cm]{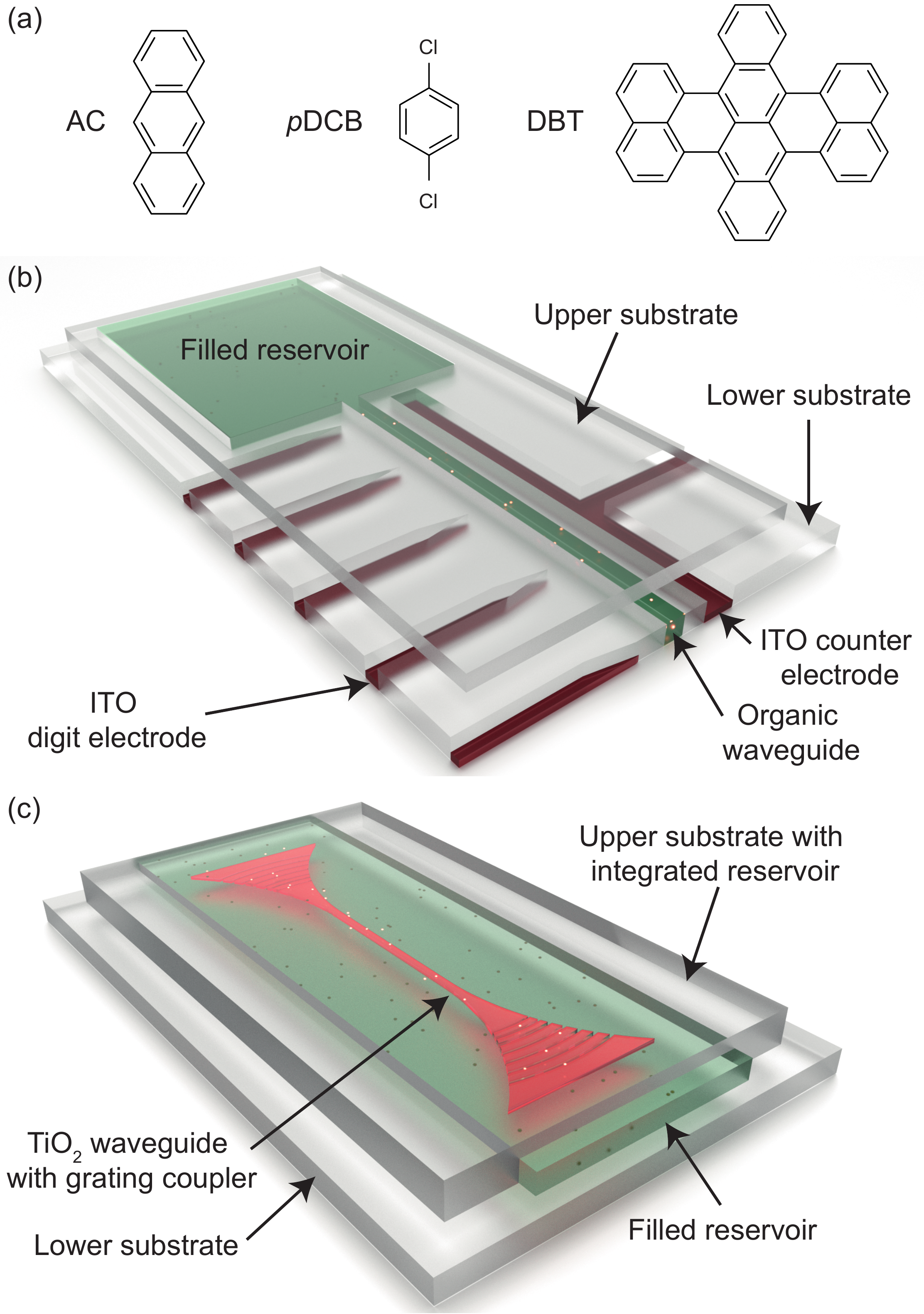}
\caption{a) Molecular structures of anthracene (AC), \textit{para}-dichlorobenzene (\textit{p}DCB) and dibenzoterrylene (DBT). The first two were used to embed DBT. b) Schematics of a nanoguide architecture, where \textit{p}DCB (green) doped with DBT molecules guides light while being surrounded by fused silica on all sides. In this structure, we also integrated indium tin oxide (ITO) microelectrodes (brown) for applying DC electric fields. The structure is cut in half along the nanoguide for illustration purposes. c) In this case, molecules embedded in AC (green) are evanescently coupled to a $\rm TiO_2$ nanoguide (red), which is terminated by integrated grating couplers. The upper substrates in (b) and (c) are offset for ease of illustration. See Suppl. Info. for fabrication details.}
\label{nanoguide-platforms}
\end{figure}

Organic dye molecules are used in a large number of applications in physics, chemistry and technology with recent important contributions to the development of fluorescence nanoscopy \cite{Hell:07}. Although the widespread dissemination of results from single-molecule biophysics has left most scientists with the impression that organic molecules photobleach and exhibit broad spectra, it turns out that polycyclic aromatic hydrocarbons (PAH) such as pentacene or dibenzoterrylene (DBT) can be nearly indefinitely photostable when embedded in organic crystals (see Fig. \ref{nanoguide-platforms}a) \cite{SMbook, Pfab:04}. Furthermore, PAHs can have very stable and narrow resonances at superfluid helium temperature, giving access to scattering cross sections close to the ideal value of $3\lambda^2/2\pi$ \cite{Gerhardt:07a}. 

The coupling between the guided mode of a nanoguide and PAHs can be achieved in two different strategies. First, one can fabricate the nanoguide from the same organic matrix that carries the PAHs, as we recently showed in a nanocapillary geometry \cite{Faez:14}. This arrangement is facilitated by the moderately high refractive index ($n$) of such matrices. In the second approach, one can place the organic matrix around a nanoguide made of material with even higher $n$ such that the PAHs would couple to its mode evanescently. In either case, it is desirable to pursue chip architectures \cite{hwang:11, Kewes:16, Lombardi:17} because compared to fiber-based systems \cite{Stiebeiner:09, Faez:14, Skoff:16}, these allow for the incorporation of micro- and nanoelectrodes to tune individual molecules to the same resonance frequency via the Stark effect and can host feedback microstructures such as photonic crystals. To this end, nanoguides on chips provide an ideal platform for scalable quantum optical networks. 

In both fabrication strategies, molecules have to be placed very close to crystal boundaries. However, one should keep in mind that PAH spectra might degrade in the close vicinity to interfaces, where the quality of the crystal is typically compromised in its last molecular layers \cite{Gmeiner:16}. This behavior is also known in other solid-state systems, where the presence of amorphous media and defects causes spectral instability of guest quantum emitters \cite{MP-special-issue-09}, making it challenging to obtain Fourier-limited spectra close to structures such as plasmonic antennas or dielectric waveguides. 

To address all the design and fabrication issues discussed above, we exploit a method based on the capillary flow of an organic matrix in the molten liquid phase and its subsequent crystallization upon controlled cooling \cite{Faez:14, Gmeiner:16}. This fabrication strategy allows us to achieve a uniform coverage of organic molecular crystals doped with DBT molecules \cite{jelezko:96} in and around dielectric nanoguides on a chip with negligible crystal defects (see Fig. \ref{nanoguide-platforms}b, c). Figure\,\ref{nanoguide-platforms}b shows the schematics of a nanoguide concept, where the organic crystal \textit{para}-dichlorobenzene (\textit{p}DCB) \cite{Verhart:16} with $n=1.54$ defines the guiding medium for light surrounded by fused silica ($n=1.46$). The low index contrast results in a low $\beta$ factor of 12\%, where $\beta$ is defined as the fraction of the power radiated by the emitter into the nanoguide mode. The advantage of this approach is that the molecules lie within the nanoguide where its radial mode profile is maximal. Figure\,\ref{nanoguide-platforms}b also sketches an arrangement of microelectrodes close to the nanoguide. Fabrication details are described in the Online Supplementary Information.

Figure\,\ref{nanoguide-platforms}c illustrates the schematics of another architecture, where the nanoguide is made of $\rm TiO_2$ ($n=2.4$) on a fused silica substrate and embedded in anthracene (AC, $n=1.7$). The two fundamental modes with orthogonal polarizations provide $\beta$ up to 28\% at the interface between the nanoguide and AC. The nanoguide cross section is adiabatically tapered on each end to a grating for facile interfacing of the guided mode with free-space beams normal to the substrate plane (see Suppl. Info).

\begin{figure}[hb]
\includegraphics[width=8.5 cm]{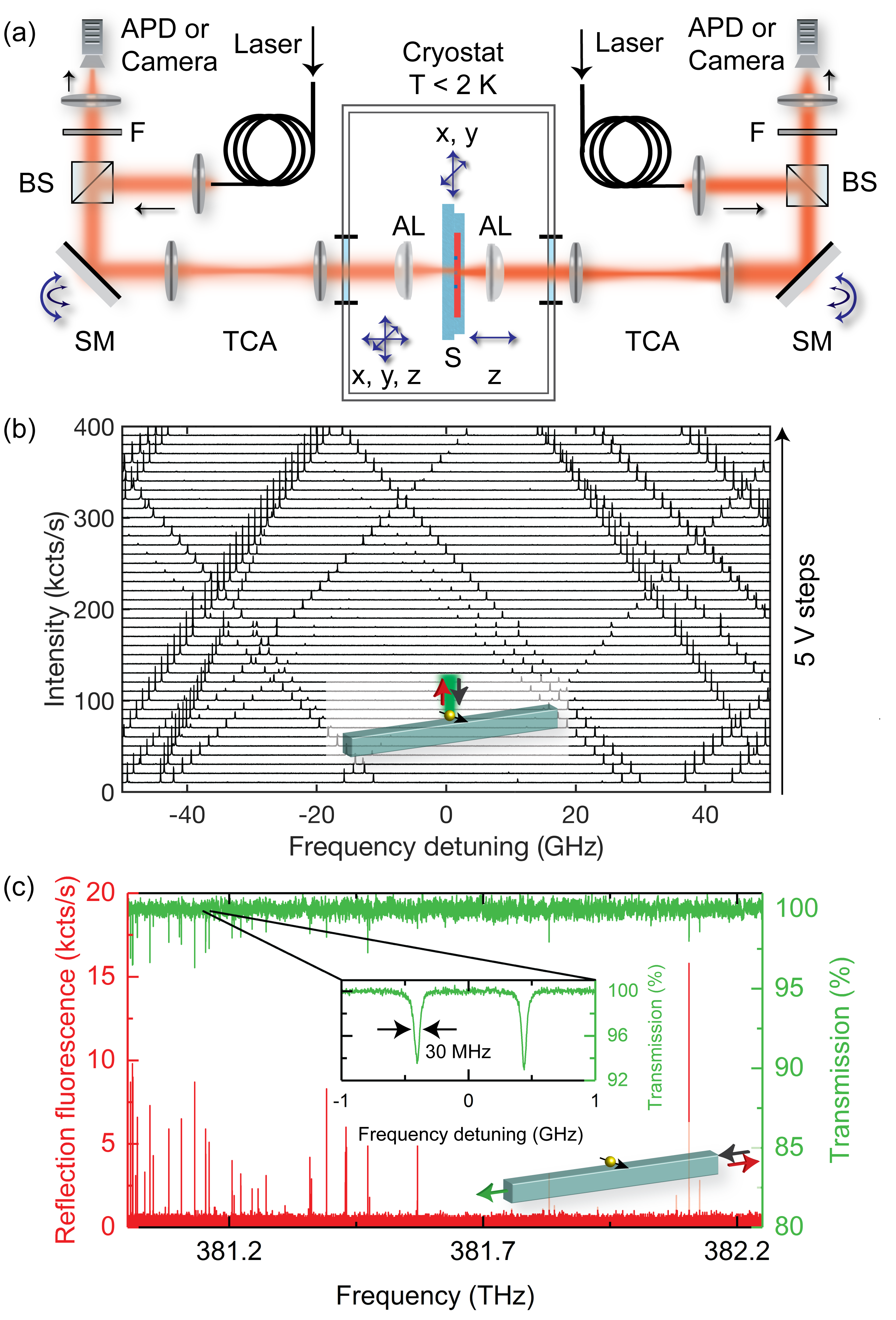}
\caption{a) Schematic of the optical setup; S: sample, TCA: telecentric lens assembly, AL: aspheric lens, SM: scanning mirror, BS: beam splitter, F: spectral filter, APD: avalanche photodiode. b) Each spectrum shows a fluorescence excitation spectrum recorded from molecules in the same excitation spot (see inset). Spectra recorded at various Stark voltages are offset in the vertical direction. c) Extinction (green) and Stokes-shifted fluorescence (red) spectra recorded through the output and input grating ports, respectively, (see inset lower right) as a function of the excitation laser frequency. The central inset shows a zoom of the extinction spectrum.}
\label{spectra} 
\end{figure}

Figure \ref{spectra}a shows the schematics of the experimental setup for spectroscopy and microscopy of single molecules coupled to nanoguides. The chip samples were mounted inside a helium bath cryostat operating at 1.7\,K and could be positioned in the substrate (x-y) plane using slip-stick piezo sliders. Two aspheric lenses with numerical aperture NA=0.77 were used to access the sample from the opposite sides, whereby one of them could be positioned in all three dimensions, and the other one was adjustable along z. The sample could be illuminated by narrow-band ($\Delta \nu < 1$\,MHz) laser beams through each arm of the resulting dual microscope. Sensitive cameras and avalanche photodiodes served as detectors. Thus, molecules could be excited and detected through the nanoguide mode or via free space at right angle. 

In cryogenic single-molecule spectroscopy, one illuminates the sample on the Fourier-limited zero-phonon line (00ZPL) of the molecules, which connects the ground vibrational levels of the electronic ground ($\left|{g, v=0}\right>$) and excited ($\left|{e, v=0}\right>$) states. Upon excitation, the upper state can decay via broad transitions to $\left|{g, v \neq 0}\right>$ levels, leading to red-shifted fluorescence. The nanoguide geometry provides convenient simultaneous access to both the incoherent red-shifted fluorescence and the coherent signal, which can be detected through the two grating ports or from the side.

Each spectrum in Fig.\,\ref{spectra}b displays fluorescence excitation spectra recorded from a few DBT molecules embedded in \textit{p}DCB within the focal spot of the laser beam (see inset). The different horizontally offset spectra were registered at different voltages applied to one of the microelectrodes shown in Fig.\,\ref{nanoguide-platforms}b, while the others were grounded. The data clearly show that the 00ZPLs of molecules within an area of diameter less than a micrometer can be tuned at about 0.5\,GHz/V over tens of GHz, which is a substantial fraction of the inhomogeneous distribution of DBT resonances \cite{Verhart:16}. We note that individual molecules respond differently to the external electric field because each experiences a different local residual matrix field \cite{Latychevskaia:02, SMbook}. Considering the very large number of molecules along the nanoguide, local Stark shifts will allow one to tune several molecules to the same frequency, realizing a mesoscopic ensemble. 

Next, we turn to the architecture of Fig.\,\ref{nanoguide-platforms}c, where molecules are evanescently coupled to the nanoguide (cross section of 200 nm $\times$ 200 nm). The green curve in Fig.\,\ref{spectra}c shows a spectrum recorded by detecting the light intensity in the nanoguide through one grating while it was excited through the other. The inset displays a high-resolution extinction spectrum, revealing sharp spectral features with Lorentzian line shapes. The observed full width at half-maximum (FWHM) of $\Gamma_0=30$\,MHz corresponds to the natural linewidth of the 00ZPL in DBT and is a robust indication for the coupling of single molecules to the nanoguide mode. Furthermore, the narrow resonances make it possible to modulate the attenuation of the light beam in the nanoguide by a small voltage applied to microelectrodes, even at MHz speeds \cite{Brunel:99}. The extinction dips in the transmission signal reach up to 7.2\%, corresponding to $\beta=7.4\%$  if we assume a combined Franck-Condon/Debye-Waller factor of 0.5 (see Suppl. Info.). 

The red spectrum in Fig.\,\ref{spectra}c presents the Stokes-shifted fluorescence recorded from the input grating port simultaneously to the extinction spectrum. While all the features in the extinction spectrum are also represented by the corresponding lines in the fluorescence signal, some resonances of the latter do not appear in the former. We attribute these to the contribution of molecules that are close to the input grating but with low coupling to the nanoguide mode. 

\begin{figure}[ht]
\includegraphics[width=8.5 cm]{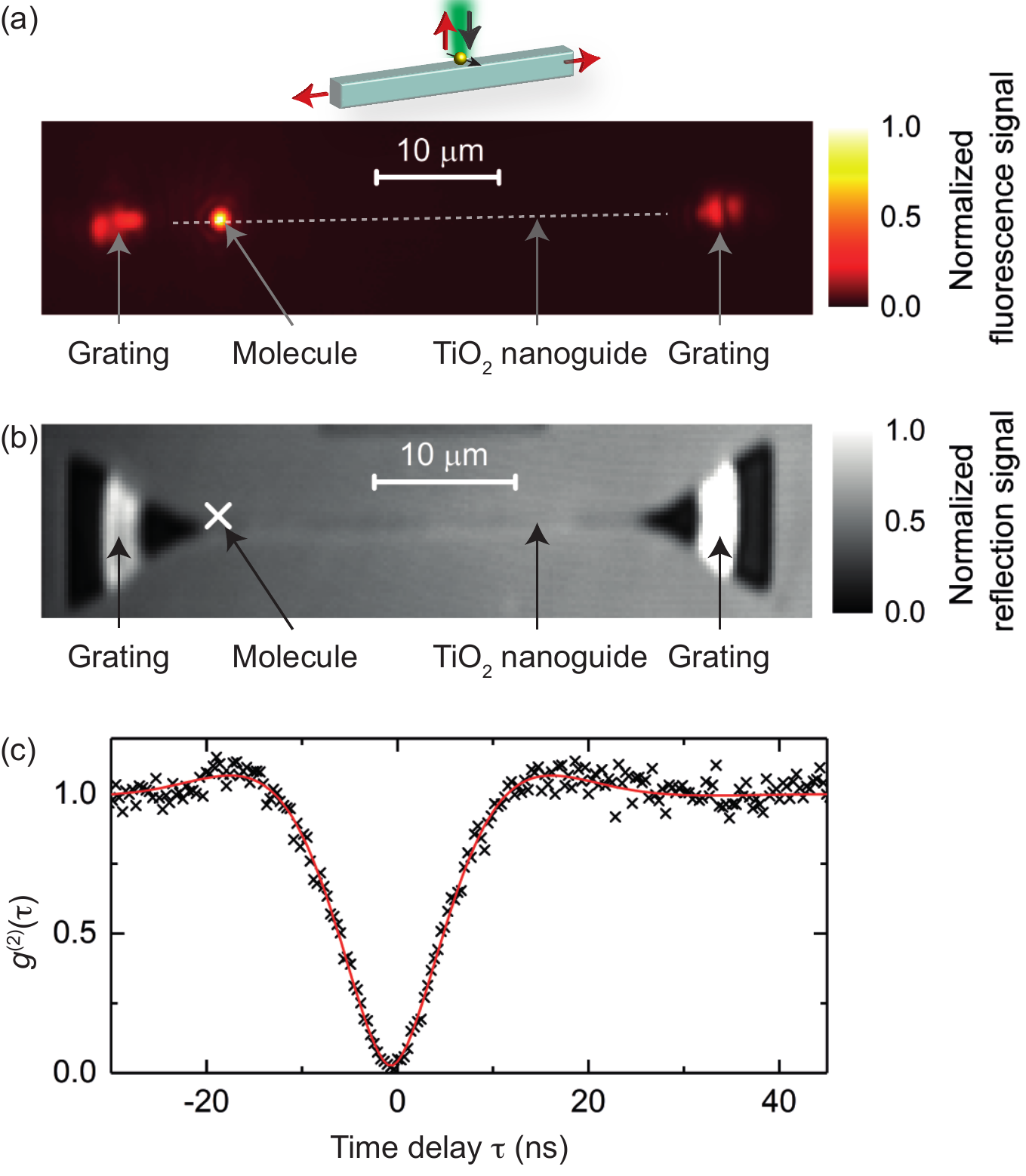}
\caption{a) Wide-field fluorescence image when a single molecule was excited by a focused laser beam from the side (see upper schematics). b) Geometry of the nanoguide structure imaged by iSCAT (see text for details). A cross shows the position of the molecule imaged in (a). c) Second-order cross-correlation measurement of the red shifted fluorescence detected through the two grating ports while the molecule was excited as in (a).}
\label{qbs}
\end{figure}

To examine the propagation of light along the nanoguide, in Fig.\,\ref{qbs}a we show a wide-field fluorescence image. Here, a single molecule was first identified in frequency space and then excited from the side in the focus of a laser beam (see inset). The three bright regions denote the Stokes-shifted fluorescence of the molecule detected directly at its location and through the two grating ports. The absence of fluorescence along the nanoguide in Fig.\,\ref{qbs}a verifies both a very low scattering loss and a high spectral selectivity to a single molecule. The image obtained via interference scattering microscopy (iSCAT) \cite{Lindfors:04} in Fig.\,\ref{qbs}b helps visualize the chip geometry. By correlating the images in (a) and (b), we found the molecule to lie at about 360\,nm from the nanoguide edge.  

Figure\,\ref{qbs}a nicely illustrates an advantage of the nanoguide architecture as an ideal integrated beam splitter for detecting the emission of a molecule in opposite directions. In this structure, we found the ratio of the fluorescence signal measured from the two gratings to yield a splitting efficiency of 57:43. By recording the two grating signals in a start-stop coincidence configuration, we obtained the second-order cross-correlation function $g^{(2)}$ shown in Fig.\,\ref{qbs}c. A very strong antibunching at zero delay confirms that the fluorescence stems from a single molecule. Moreover, the theoretical fit at a Rabi frequency of 0.9$\Gamma_0$ lets us determine an excited-state lifetime of 5\,ns. We remark in passing that cross-correlation measurements in a one-dimensional nanoguide provide an ideal platform for revisiting the particle nature of spontaneous emission since a detector click either occurs on the right or the left channel \cite{Grangier:86}. 

\begin{figure}[t!]
\includegraphics[width=8.5 cm]{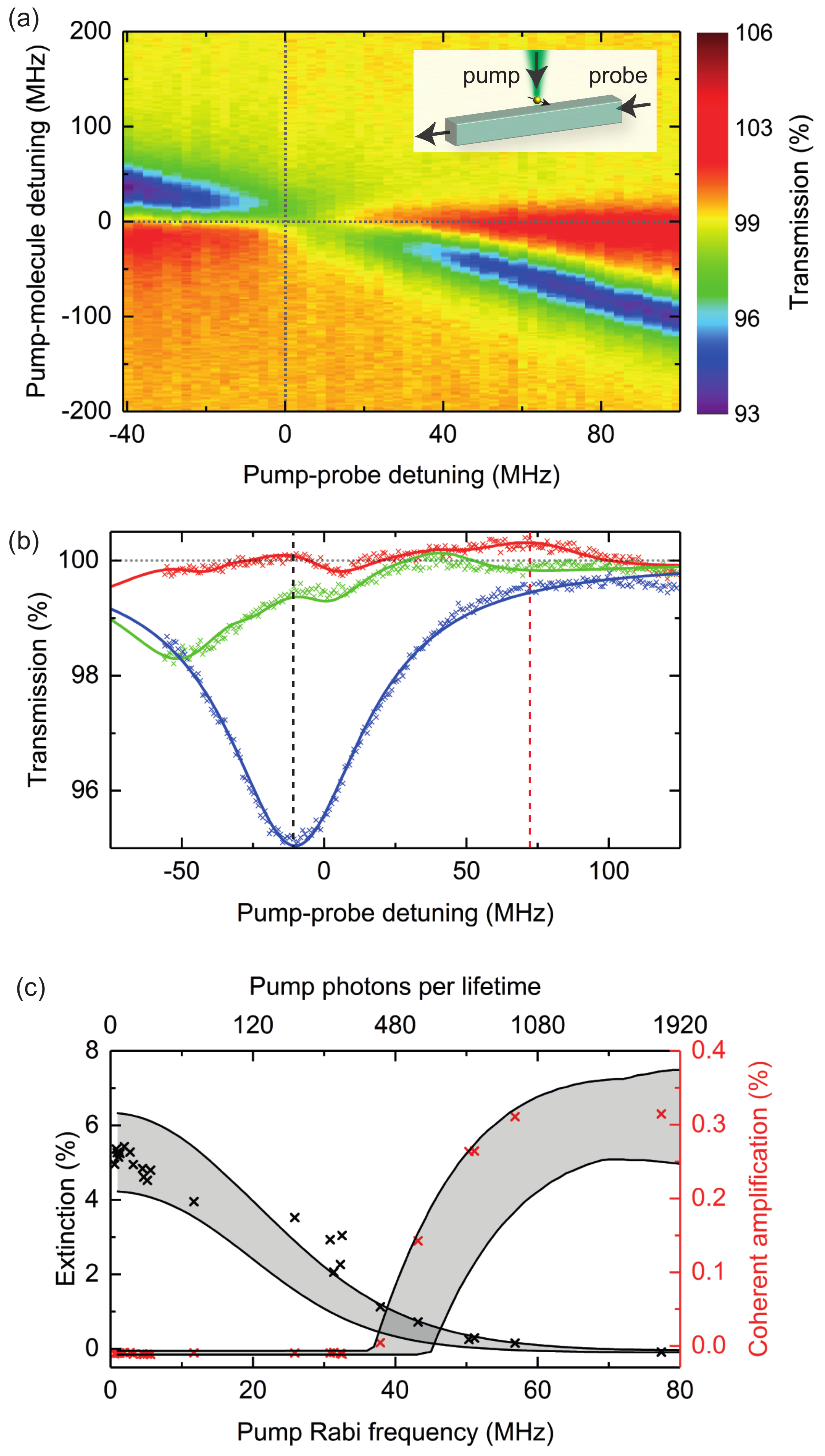}
\caption{a) Nanoguide transmission as a function of pump-molecule and pump-probe frequency detunings for fixed probe and pump Rabi frequencies of $\Omega_{\rm prb}=0.1 \Gamma_0$ and $\Omega_{\rm pmp}=\Gamma_0$, respectively. b) Exemplary nanoguide transmission spectra for three different pump intensities $\Omega_{\rm pmp}$=0.5 MHz (blue), $\Omega_{\rm pmp}$=43 MHz (green), $\Omega_{\rm pmp}$=77 MHz (red). The vertical dashed lines point to the strongest attenuations and amplifications of the probe beam. c) Extinction signal (black) and coherent amplification (red) of the probe beam as a function of Rabi frequency and incoming photon number of the pump beam. The grey regions correspond to expected variations caused by a very small spectral diffusion less than $0.2\Gamma_0$}
\label{nonlinear}
\end{figure}

A single PAH molecule can act as a highly nonlinear medium for single photons if it is efficiently coupled to a propagating mode. Such a nonlinearity can be exploited to switch the propagation of photons by using an external control laser field which dresses the molecular electronic states \cite{Maser:16}. Our current nanoguide geometry provides a particularly convenient configuration for such studies because by focusing the stronger pump beam from the side and detecting the weaker probe beam through the nanoguide, one can easily separate the two in the detection channel (see inset in Fig.\,\ref{nonlinear}a). Figure\,\ref{nonlinear}a shows a map of the nanoguide transmission for various frequency detunings between the pump laser and the molecule as well as between the pump and probe lasers. Here, the pump strength was chosen to correspond to a moderate Rabi frequency of $\Omega_{\rm pmp}=\Gamma_0$ while the probe was kept weak at $\Omega_{\rm prb}=\Gamma_0/10$. The map displays regions where the probe signal is attenuated by about 7\% and other parts where one detects about 5\% more signal than the probe intensity. It is important to note, however, that the latter is predominantly due to the direct resonant scattering of the pump beam by the molecule into the nanoguide mode. To assess the amplification of the probe beam, we normalized the signal to its value far from the molecular resonance while keeping the pump-molecule frequency detuning constant. In Fig.\,\ref{nonlinear}b, we display examples of the probe transmission versus pump-probe detuning for three different pump Rabi frequencies of $0.02\Gamma_0$ (blue), $1.4\Gamma_0$ (green), and $2.6\Gamma_0$ (red). The vertical dashed lines point to two frequency regions, where the probe attenuation and amplification are switched most sensitively as the pump power is increased. We find the net coherent amplification of the probe beam to reach 0.3\%. 

The symbols in Fig.\,\ref{nonlinear}c summarize the degrees of probe extinction and amplification measured for many pump powers. The shadowed regions of the solid curves show the corresponding calculated quantities, allowing for a small spectral instability less than $0.2\Gamma_0$, which occurred over several hours during these measurements (see Suppl. Info.). These data show that in addition to a static Stark effect discussed above, one can also use an external optical field to control the propagation of a light beam in the nanoguide. The latter option is particularly attractive because it can be implemented with a diffraction-limited spatial resolution and at speeds up to hundreds of MHz.

In this Letter, we presented linear and nonlinear measurements on individual molecules coupled to waveguides on a chip and showed that the propagation of light in a nanoguide can be switched by external electrostatic or optical fields. The demonstration that nanofabricated structures can be overlapped with molecular crystals over large areas opens realistic prospects for investigating polaritonic and quantum many-body phenomena while controlling the contribution of each single emitter. Future employment of optical circuit elements such as beam splitters, routers and interferometers \cite{Lipson:06} together with novel devices such as on-chip superconducting detectors \cite{Akhlaghi:15} will also allow the realization of integrated quantum circuits. A particular advantage of chip architectures is their ease of interfacing with other chip-based materials such as color centers \cite{Sipahigil:16} or rare-earth ions \cite{Utikal:14, Marzban:15} to realize hybrid quantum systems. 

We thank Harald Haakh for fruitful discussions and help with theoretical design and Maksim Schwab for his contributions to the mechanical components. This work was supported by an Alexander von Humboldt professorship, the Max Planck Society and the European Research Council (Advanced Grant SINGLEION).


%

\newpage
\end{document}